\documentclass[aps,prl,superscriptaddress,showpacs,twocolumn]{revtex4}
\usepackage{graphicx}
\pdfoutput=1
\begin{document}
%
\title{Bistability in superconducting rings containing an
inhomogeneous Josephson junction}
%
\author{M.~Gaass}
\affiliation{Institute for experimental and applied Physics,
University of Regensburg, 93025 Regensburg, Germany}
\author{S.~Nadj-Perge}
\affiliation{Department of Physics, University of Belgrade, P.O. Box
368, 11001 Belgrade, Serbia}
\author{Z.~Radovi\'{c}}
\affiliation{Department of Physics, University of Belgrade, P.O. Box
368, 11001 Belgrade, Serbia}
\author{A.~Bauer}
\affiliation{Institute for experimental and applied Physics,
University of Regensburg, 93025 Regensburg, Germany}
\author{M.~Aprili}
\affiliation{Laboratoire de Physique des Solides, Univ.~Paris-Sud,
CNRS, UMR 8502, 91405 Orsay Cedex, France} 
\author{W.~Wegscheider}
\affiliation{Institute for experimental and applied Physics,
University of Regensburg, 93025 Regensburg, Germany}%
\author{C.~Strunk}
\affiliation{Institute for experimental and applied Physics,
University of Regensburg, 93025 Regensburg, Germany}%

%
%
\begin{abstract}
We investigate the magnetic response of a superconducting Nb ring
containing a ferromagnetic PdNi Josephson junction and a tunnel
junction in parallel. A doubling of the switching frequency is
observed within certain intervals of the external magnetic field.
Assuming sinusoidal current-phase relations of both junctions our
model of a dc-SQUID embedded within a superconducting ring explains
this feature by a sequence of current reversals in the ferromagnetic
section of the junction in these field intervals. The switching
anomalies are induced by the coupling between the magnetic fluxes in
the two superconducting loops.

%
\end{abstract}
%
%
%
%

%
\maketitle


A Superconducting Quantum Interference Device (SQUID) consists of a
superconducting loop interrupted by one (rf SQUID), or two (dc
SQUID) Josephson junctions. The magnetic moment of rf SQUIDs and the
critical current of dc SQUIDs are periodic functions of the magnetic
flux $\Phi$ enclosed by the loop. In most cases, the periodicity is
given by the magnetic flux quantum $\Phi_0=h/2e$. This is a
consequence of the gauge-invariant connection between $\Phi$ and the
phase difference $\varphi$ across the junction(s) via the
current-phase relation (CPR) $I_s(\varphi)$. Under certain
conditions, the CPR of the junction(s) is not a simple sine, but can
contain higher harmonics: $I_s(\varphi)=\sum_n\;I^{(n)}_c\sin
n\varphi$, where the coupling coefficients $I^{(n)}_c$ quantify the
relative strength of processes with a coherent transfer of $n$
Cooper-pairs \cite{golubov}. In some cases, the first order
coefficient $I^{(1)}_c$ can even vanish, for instance, in asymmetric
$45$ degree grain boundary junctions in d-wave symmetry
superconductors \cite{mannhart,lindstrom}, in out of equilibrium SNS
junctions \cite{baselmans} and for ballistic SFS junctions at the
$0$-$\pi$ transition \cite{radovic1,radovic2}. The remaining second
order coefficient $I_c^{(2)}$ will result in a doubling of the
frequency in the interference pattern. Although, a frequency
doubling has been observed experimentally \cite{sellier}, its origin
is still under debate, since there are also dynamic effects in
inhomogeneous junctions, which can lead to this effect
\cite{frolov2}.

Recently, it was observed that the critical current of high
temperature superconductor dc-SQUIDs shows a $\Phi_0/2$ periodicity
in certain sections of the interference pattern \cite{lindstrom}. To
explain this observation it was suggested that the random faceting
of the grain-boundary induces a distribution of 0 and $\pi$
couplings along the junction. For some parti\-cular values of the
applied magnetic field the first order coupling of the overall
junction is zero, allowing second order coupling to be dominant. A
similar effect has been also measured in the multi terminal
transport of a 2-dimensional electron gas connected to a
superconducting loop \cite{klapwijk}.

Here we report on a different material system, showing a similar
phenomenology and suggest an alternative interpretation of this
phenomenon. We study the magnetic response of an rf SQUID where the
junction is inhomogeneous and formed by the parallel connection of a
conventional ($0$) Josephson junction and a ferromagnetic $\pi$
junction. By measuring the total flux in the SQUID while increasing
the external magnetic field $B_{ext}$, we mostly observe a
$\Phi_0$-periodic penetration of flux quanta into the loop, every
time the critical current of the junction is exceeded. However, for
some values of the external magnetic field we find a doubling of the
switching frequency. The proposed model explains this effect in
terms of a bistability of the supercurrent in the $\pi$-junction for
certain values of the applied magnetic field. Our double SQUID model
successfully explains the magnetic field and temperature dependence
of our observations, although the CPR of both junctions is assumed
to be purely sinusoidal.

\begin{figure}
\includegraphics[width=7cm, keepaspectratio]{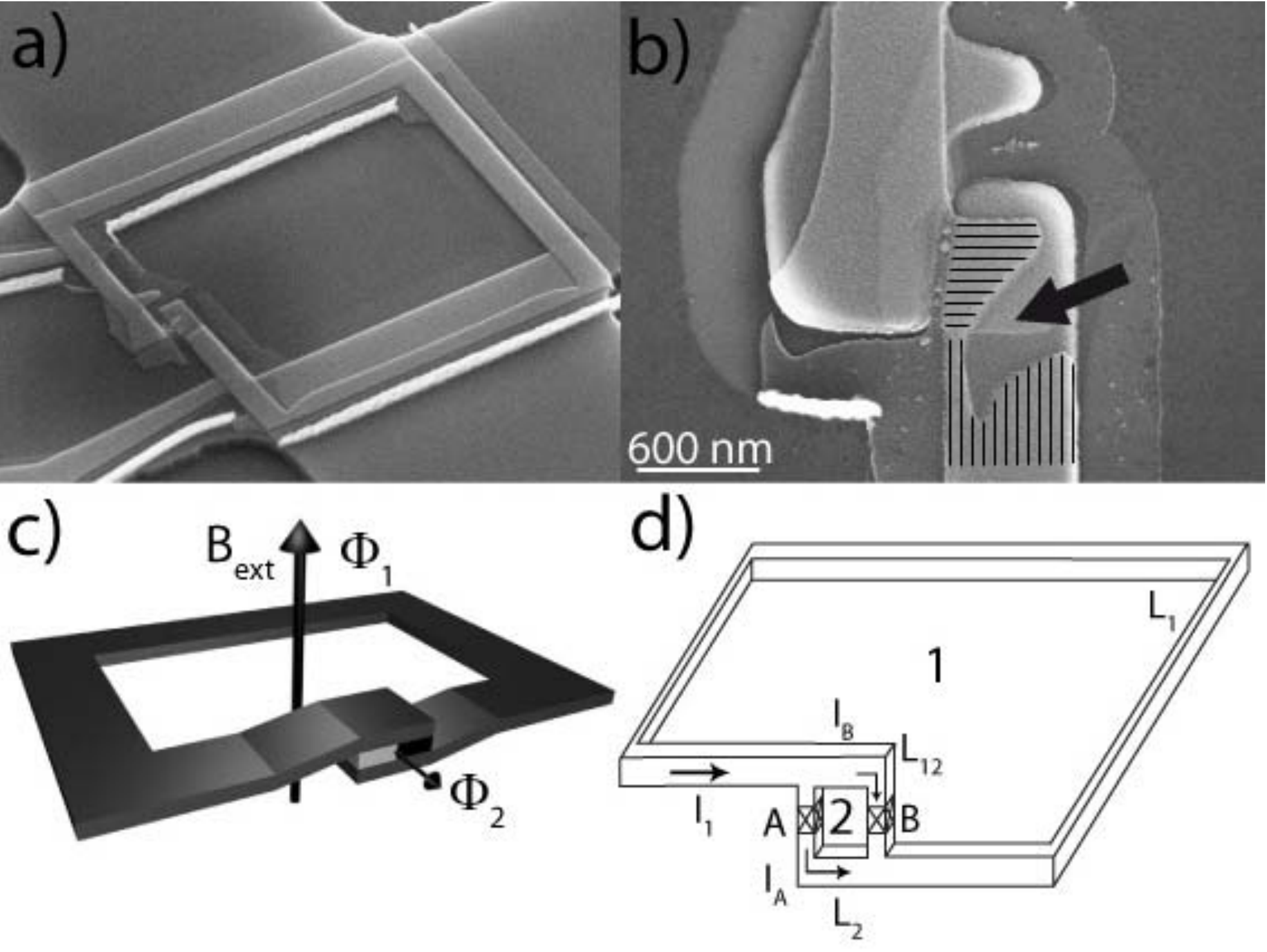}
\caption{(a) Scanning electron image of the loop on top of the
active area of the Hall-sensor. Two leads connected on both sides of
the junction are used to control the flux through the loop. (b) A
zoom onto the junction area showing the parallel connection of the
Nb/Nb-contact (junction A, vertical lines) and the Nb/PdNi/Nb-
(junction B, horizontal lines). (c) Schematic of the device
geometry. (d) Schematic of the equivalent double SQUID: Loops 1 and
2 with self-inductances $L_1$, $L_2$ and mutual inductance $L_{12}$
containing junctions A and B.} \label{sample_overview}
\end{figure}

We use Nb as the superconductor and dilute PdNi as the ferromagnet
for our SFS junctions. To pattern the loops, we use a robust
Si$_3$N$_4$/PES mask system for the shadow evaporation. The PES
(polyether sulfone) forms a highly thermostable sacrificial layer
\cite{dubos}. The 60 nm thick Si$_3$N$_4$ was deposited by plasma
enhanced chemical vapor deposition on top of the PES and provides
sufficient mechanical stability to resist the large stresses created
by the Nb film. After patterning the mask by electron beam
lithography and reactive ion etching using CHF$_3$, the Si$_3$N$_4$
mask was underetched by an isotropic oxygen-plasma. The undercut can
have a value of up to 1~$\mu$m. Evaporation of 40~nm of Nb and 10~nm
of PdNi under different angles provides the desired superconducting
loops with an integrated SFS planar junction, as illustrated by a
scanning electron micrographs of a sample in
Figs.~\ref{sample_overview}a and b. In Figs.~\ref{sample_overview}c
and d the equivalent schematics are shown.
 The thickness of the PdNi film was chosen to produce a $\pi$ junction
close to the 0-$\pi$ crossover \cite{kontos1}. A 10\% misalignment
of the sample during evaporation of the second Nb-layer resulted in
an overlap of the two Nb-films without PdNi-interlayer, as indicated
by the arrow in Fig.~\ref{sample_overview}b. Strong gettering of
residual gas by the Nb during evaporation of PdNi results in a
rather transparent tunneling contact between the two Nb-films in
this area, as a second Josephson junction.

We have placed the sample on top of the active area of a
micron-sized Hall-sensor in order to detect the magnetic response
\cite{bauer,geim}. The Hall-sensor is realized in a semiconductor
heterostructure having the electron density of $2.25 \times
10^{15}\,\rm{m}^{-2}$ and the mobility of $1.13 \times
10^{6}\,\rm{cm}^2/\rm{Vs}$. We achieve a sensitivity of roughly
$500\,\rm{nT}/\rm{Hz}^{1/2}$ to $50\,\rm{nT}/\rm{Hz}^{1/2}$
depending on the sensor current. For our loop dimensions of
approximately $7.6\,\rm{\mu m} \times 8.5\,\rm{\mu m}$, the magnetic
flux quantum $\Phi_0 = 2.067 \times 10^{-15}\,\rm{Vs}$ corresponds
to a magnetic field of about $31\,\rm{\mu T}$.

\begin{figure}
\includegraphics[width=7cm, keepaspectratio]{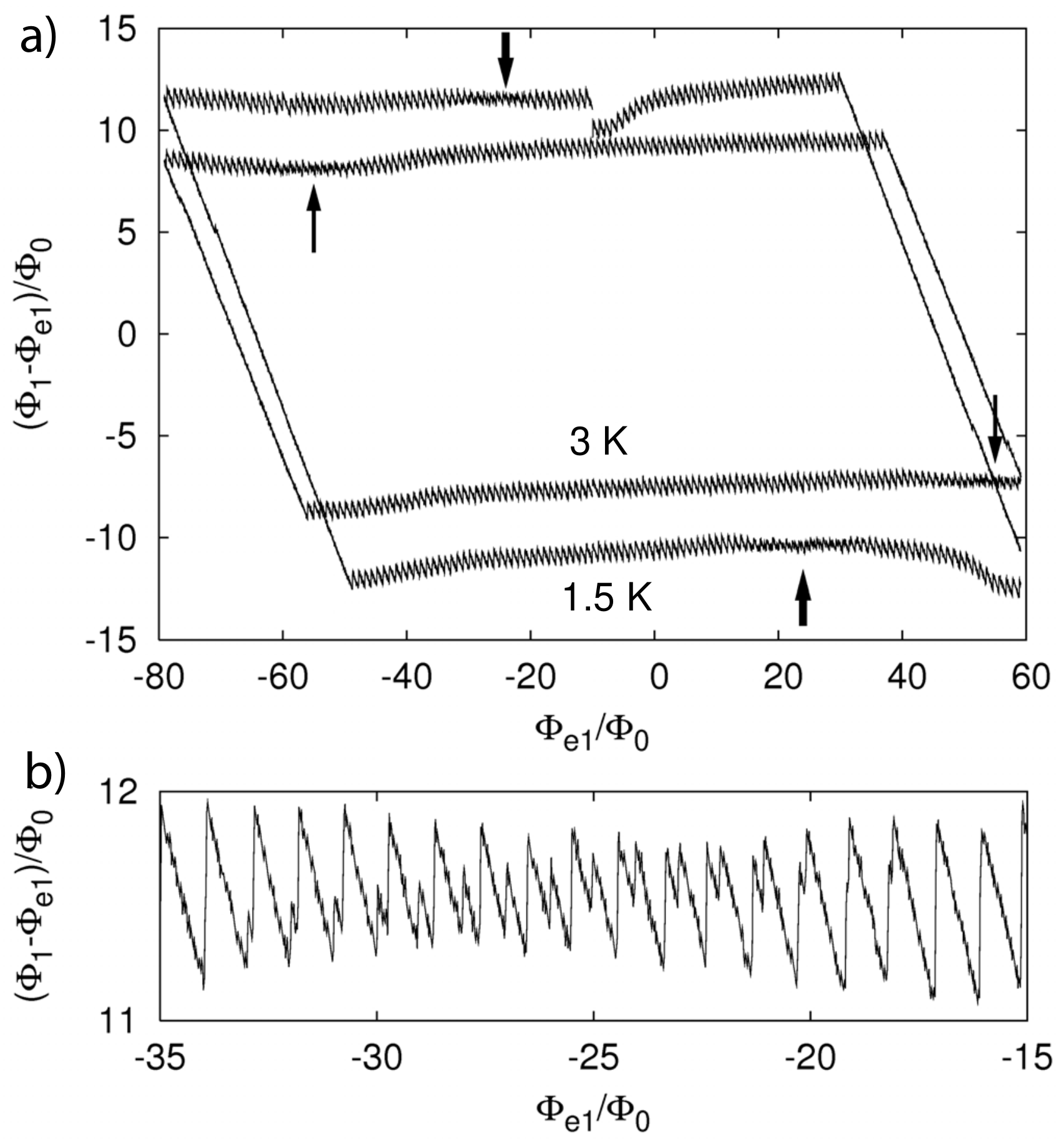}
\caption{(a) Full hysteresis cycle for two different $T=1.5$~K
(outer cycle) and 3.0~K (inner cycle), corresponding to the
critical currents $I_{c1}\simeq 980$ and $690$ $\mu A$
respectively. The arrows indicate positions of substructures with
double frequency. With increasing T the substructures are shifted
to higher fields. The ring inductance $L_1\simeq 26$~pH is
determined from the estimated filling factor of the magnetometer.
The irregular drifts and jumps are artifacts of the Hall cross.
(b) Magnified top substructure on outer cycle is shown as an
example (All other substructures are similar). The absolute value
of the induced flux $\Phi_1-\Phi_{e1}$ is determined from the jump
height with an uncertainty of $0.1\Phi_0$. }\label{doubling}
\end{figure}

Upon sweeping the external magnetic field,  circulating
supercurrents in the loops are generated. The resulting flux in the
loop, $\Phi_1$, induces a flux-periodic contribution to the Hall
voltage across the Hall sensor, while the contribution of the
external flux $\Phi_{e1}$ is simply subtracted. A typical trace of
the induced flux $L_1I_1=\Phi_1-\Phi_{e1}$ vs. $\Phi_e$ is depicted
for two different temperatures in Fig.~\ref{doubling}a. Only the
flux through the ring is measured due to negligible contribution of
the almost orthogonally tilted micro loop, see
Fig.~\ref{sample_overview}c. The magnetic response of the ring is
strongly hysteretic due to the large $LI_c$-product, the latter is
characterized by the parameter $\beta_{L1}=2\pi L_1I_{c1}/\Phi_0\gg
1$ \cite{barone}. The ring inductance is determined from the
estimated filling factor of the magnetometer. The critical current
is determined from the vertical size of the hysteresis loops in
Fig.~\ref{doubling}a.

The signal shows additional substructures in the switching pattern
in certain intervals of the external magnetic field, as indicated by
the arrows in Fig.~\ref{doubling}a. This effect has been seen in
three samples with similar inhomogeneous junction geometry. A zoom
onto the top substructure in the outer cycle is shown in
Fig.~\ref{doubling}b. All other substructures are similar. It can be
seen that the regular $\Phi_0$-periodic switching pattern is
interrupted by additional peaks, which gain height, until they take
over. The field intervals displaying the substructure in the
switching behavior shift towards higher field when the temperature
is increased. The substructure looks similar to the predicted
frequency doubling effect, expected for a dominant second harmonic
contribution to the CPR \cite{radovic2}. However, as we show below,
the observed substructure can be explained in terms of two coupled
loops, even for a sinusoidal CPR of both $0$ and $\pi$ junctions.

In the following model treatment, we approximate the extended
inhomogeneous junction by a parallel connection of two short
junctions A and B, the latter being in the $\pi$-state. We assume a
sinusoidal current-phase relations
$I_{A(B)}=I_{cA(B)}\sin\varphi_{A(B)}$. These two junctions form a
small dc SQUID with an inductance $L_2$, which interrupts the large
loop with inductance $L_1$. As sketched in
Fig.~\ref{sample_overview}d, our model system then consists of a rf
SQUID, with an embedded dc SQUID as weak link. The supercurrents
circulating in the two SQUID loops are coupled by the Kirchhoff laws
and their mutual inductance $L_{12}$.

The free energy of the circuit is given by
\begin{equation}
W=E_A(\varphi_A)+E_B(\varphi_B)+\frac{1}{2}L_1I_1^2+\frac{1}{2}L_2I_2^2+L_{12}I_1I_2 \label{U}
\end{equation}
where $\varphi_A$ and $\varphi_B$ are the macroscopic phase
differences across the junctions A and B, $I_1=I_A+I_B$ is the
current circulating in the large loop, $I_2=I_A$ is the current
circulating in the small loop, where $I_A$ and $I_B$ are the
currents through junctions A and B, respectively. The first two
terms in Eq.(\ref{U}) are the Josephson energies
$E_i(\varphi_i)=(\Phi_0/2\pi)|I_{ci}|(1\mp\cos\varphi_i),\;(i=A,B)$
for 0- and $\pi$-junctions, respectively. The three remaining terms
represent the magneto-static energy \cite{barone,landau}. The
magnetic fluxes $\Phi_1$ and $\Phi_2$ through the loops 1 and 2 are
given by
\begin{equation}
\Phi_1=\Phi_{e1}-L_1I_1-L_{12}I_2, \;\;
\Phi_2=\Phi_{e2}-L_{12}I_1-L_{2}I_2. \label{phi}
\end{equation}
Here, $\Phi_{e1}$ and $\Phi_{e2}$ are the corresponding fluxes of
the external magnetic field. The total fluxes are related to the
phase differences
\begin{equation}
\Phi_1=\frac{\Phi_0}{2\pi}\varphi_B,\quad
\Phi_2=\frac{\Phi_0}{2\pi}(\varphi_A-\varphi_B).
\end{equation}
Finally, Eq.(\ref{U}) can be rewritten explicitly as a function of
$\Phi_1$ and $\Phi_2$ in the form
\begin{eqnarray}
W&=&E_A(\varphi_A)+E_B(\varphi_B)\nonumber \\
&+&\frac{1}{L_1L_2-L_{12}^2}\Big\{\frac{L_2}{2}(\Phi_1-\Phi_{e1})^2+
\frac{L_1}{2}(\Phi_2-\Phi_{e2})^2 \nonumber\\
 &-&L_{12}(\Phi_1-\Phi_{e1})(\Phi_2-\Phi_{e2})\Big\}.
\end{eqnarray}
\begin{figure}[t]
\includegraphics[width=8cm,angle=0, keepaspectratio]{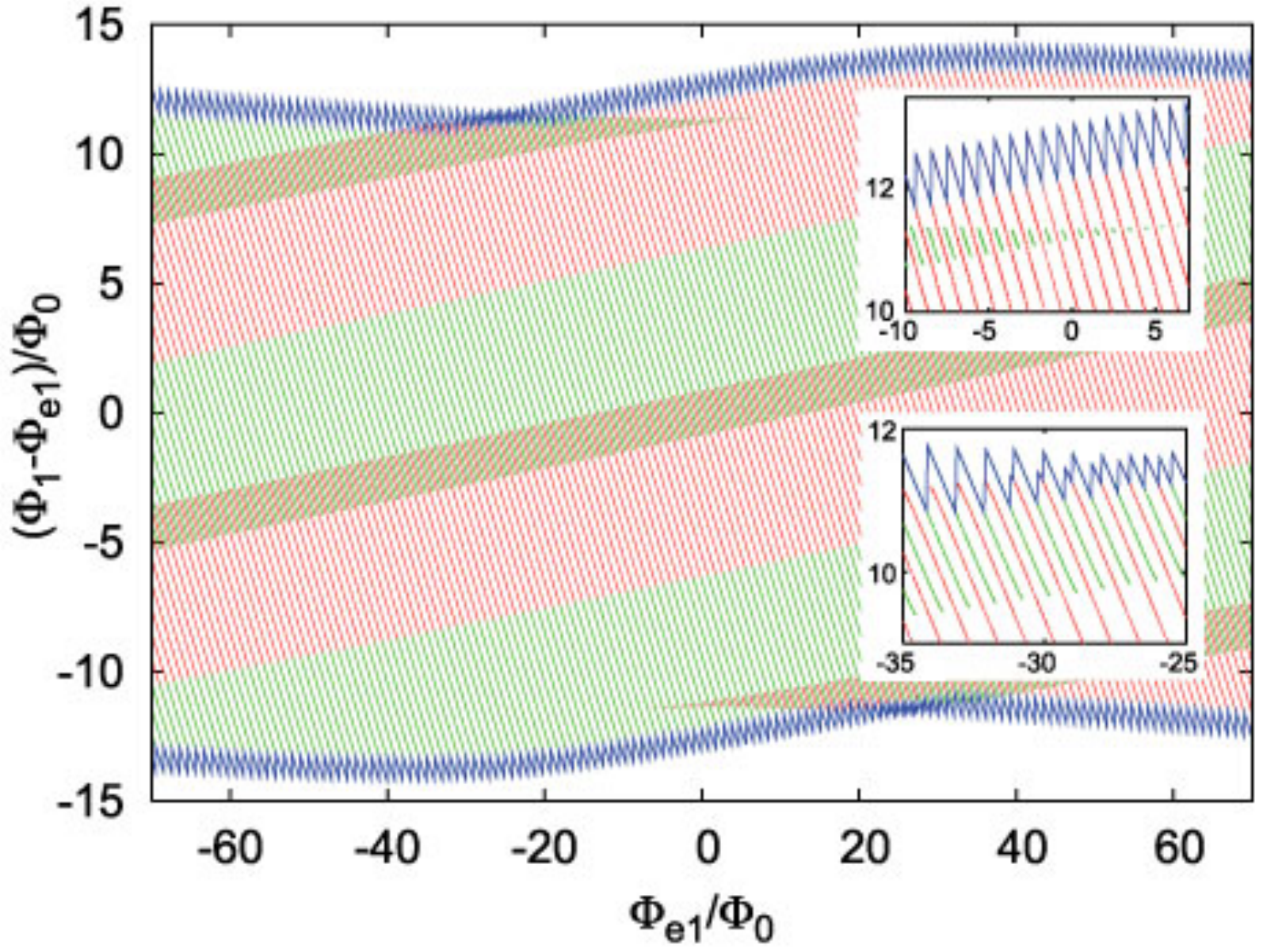}
\caption{The map of local energy minima of the double SQUID. Red and
green lines correspond to positive and negative sign of the current
$I_B$. Top and bottom sawtooth lines represent the largest
hysteresis in $\Phi_1-\Phi_{e1}$. Parameters are the same as in
Fig.~\ref{doubling_theory} (outer cycle). Upper inset: Nucleation of
states with opposite $I_B$(green). Lower inset: Alternation of
states (red and green lines) with positive and negative sign of
$I_B$ in the region with frequency doubling.} \label{free_ene}
\end{figure}

For given external fluxes $\Phi_{e1}$ and $\Phi_{e2}$, the local
minima of the free energy $W$ with respect to $\Phi_1$ and $\Phi_2$
are calculated numerically and plotted in Fig.~\ref{free_ene}. Like
for usual rf SQUIDs in the hysteretic regime, the circulating
current in the large loop is a multivalued function of external flux
if $\beta_{L1}>1$. When the external magnetic field is swept, e.g.,
in negative direction, the circulating supercurrent follows the
lines of local energy minima, as indicated by the red lines in
Fig.~\ref{free_ene}. At the upper end of each line the state becomes
unstable and the system switches into the nearest available state
with lower free energy, as reflected by the sharp drops in $\Phi_1$.
The envelope of these jumps is periodic, due to a modulation of the
maximum $I_1$ by the integrated dc SQUID. This comes from the
coupling between $\Phi_1$ and $\Phi_2$, implying a circulating
current also in the small loop (e.g. with $I_B>0$), which
contributes to the free energy. Around $\Phi_{e1}/\Phi_0\approx5$,
states with opposite current $I_B$ in the small junction become
stable (see the short green lines in the upper inset in
Fig.~\ref{free_ene}). The stability region for this set of states
grows until they eventually become more stable than those with the
original (positive) orientation of $I_B$. This is precisely the
region, where these states become involved in the switching process.
In this region, states with positive and negative orientation of the
current in the small loop alternate, resulting in a doubling of the
switching frequency (lower inset in Fig.~\ref{free_ene}). When the
external flux is decreased further, the states with the positive
orientation of $I_B$ become energetically unfavorable and their
region of stability shrinks, until the switching processes entirely
involve states with negative $I_B$ (green lines). In this way, the
doubling of the switching frequency is traced back not to a period
doubling in the CPR, but to the presence of the two-fold orientation
of the current in the $\pi$-section of the junction.

\begin{figure}[t]
\includegraphics[width=7cm, angle=0, keepaspectratio]{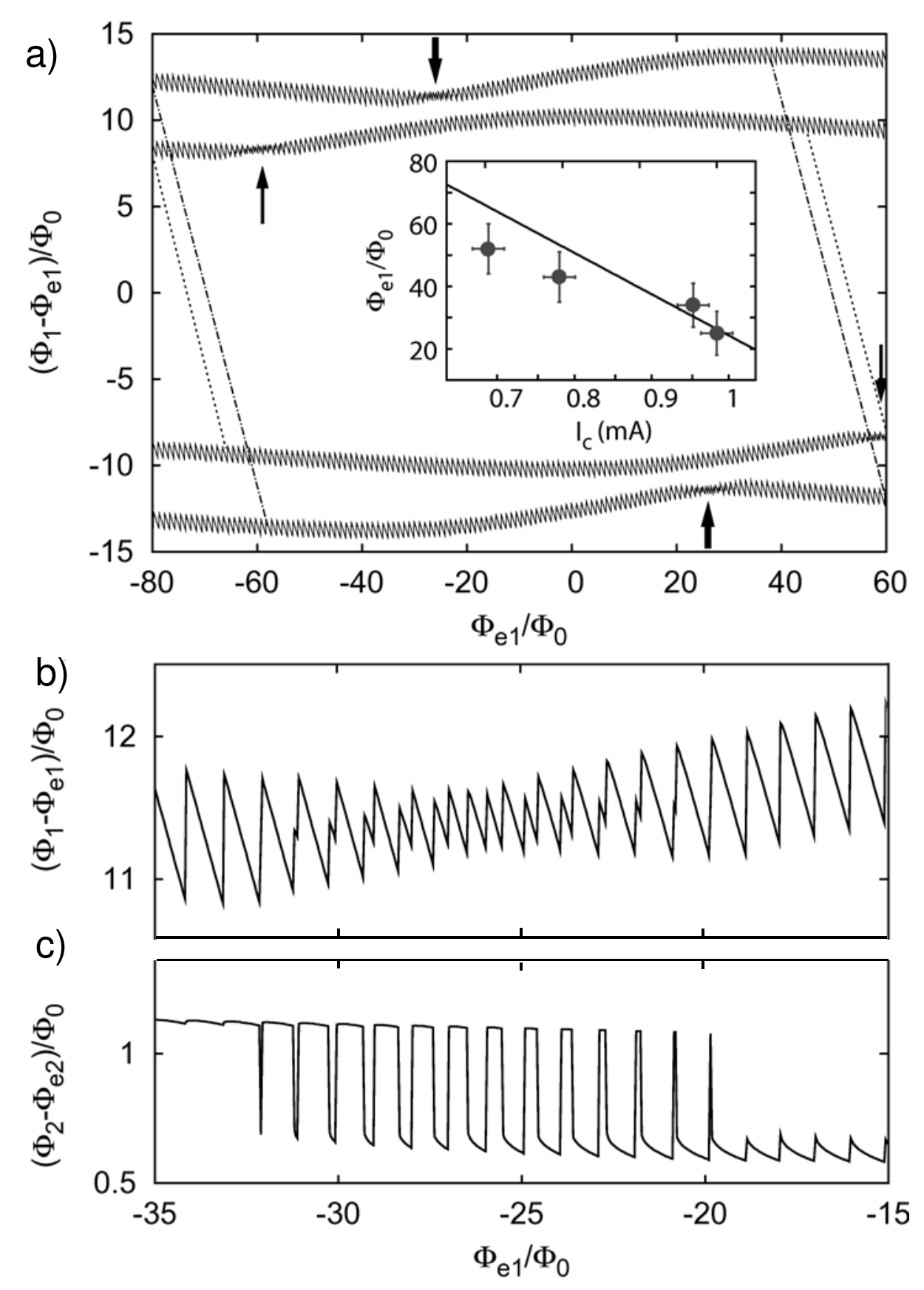}
 \caption{(a) Calculated largest hysteretic cycles fitted to the
experimental data shown in Fig \ref{doubling} with the parameters
given in the text. The inner cycle corresponds to a reduction of
$I_{cA}$ and $I_{cB}$ by $25\%$, corresponding to the higher
temperature. The dotted lines correspond to the reversal of the
sweep direction in the experiment. Inset: Shift of the switching
anomaly with the measured total critical current $I_C(T)$ at
temperatures 1.5, 2, 2.5 and 3 K (dots). The solid line represents
the theoretical prediction with the same set of parameters (b) Zoom
on the upper substructure in the outer cycle. (c) Corresponding
induced flux through the small loop.} \label{doubling_theory}
\end{figure}

To compare the numerical results with our experiment we take
junction A in the $0$ state (tunnel junction, $I_{cA}>0$), junction
B  in the $\pi$ state
 (SFS, $I_{cB}<0$),
and $I_{cA}/I_{cB}=-10$, $\beta_{L1}=90$, $\beta_{L2}= 2\pi
L_2I_{cB}/\Phi_0=-1.5$, $2\pi L_{12} I_{cA}/\Phi_0=-9.7$ and
$\Phi_{e2}/\Phi_{e1}=-0.005$. The critical current $I_{cA}$ and
inductance of the ring $L_{1}$ are determined directly from the
experimental hysteresis loop in Fig.~\ref{doubling}a, while $I_{cB}$
is estimated from previous measurements on similar Nb-PdNi junctions
\cite{bauer}. The dc SQUID inductance $L_{2}$ and mutual inductance
$L_{12}$ affect mainly size and position of the switching anomaly on
the hysteresis loop. For reasonable values of chosen parameters a
good agreement is achieved between the experimental data for
$T=1.5$K and numerical simulations (see Figs. \ref{doubling} and
\ref{doubling_theory}). Note that the ratio $\Phi_{e2}/\Phi_{e1}$ of
external fluxes differs from the loops area ratio and has (for this
sample) negative sign due to the almost orthogonal tilt of the small
loop with respect to the sample. Its value determines the number of
the observed double switching events and it can be readily
determined from the data.

If we assume that the measured reduction of $I_{cA}$ to
$\approx$~75\% at $T=3$~K (see Fig.~\ref{doubling_theory}) is
similar also in $|I_{cB}|$, we can well reproduce the observed shift
of the switching anomaly towards higher external flux (inner trace
in Fig.~\ref{doubling_theory}a). The dots in the inset in
Fig.~\ref{doubling_theory}a show the shift of the switching anomaly
vs.~the critical current $I_C(T)$ for different temperatures. The
solid line denotes the prediction of our theory using the same set
of parameters.

Sequences of the frequency doubling in the flux modulation
periodically occur in certain intervals of the external magnetic
field, if $|\beta_{L2}|\gtrsim 1$ of the smaller loop is
sufficiently large.  A zoom onto the substructure is shown in Fig.
\ref{doubling_theory}b. Substructures in the switching pattern
$\Phi_1-\Phi_{e1}$ occur with the period $|\Phi_{e1}/\Phi_{e2}|$.
The corresponding flux modulation in the small loop is shown in Fig.
\ref{doubling_theory}c. When the external field is varied, small
periodic perturbations induced by the large loop alternate
periodically the energies of the two opposite current directions in
the small loop, thus forming the bistable region with large flux
oscillations. The bistable region is placed symmetrically around the
external flux value corresponding approximately to integer number of
$\Phi_0$ in the small loop and equal energies for opposite currents
in junction B.

Additional calculations show that in the ground state the first
substructure occurs in the low external field corresponding to
$\Phi_{e2}\approx0$ \cite{stevan}. When both junctions are in the
$0$ state the first substructure is located at
$\Phi_{e2}\approx\Phi_0/2$, which corresponds in our case to a very
large $\Phi_{e1}$. However, positions of substructures on the
hysteretic loop are strongly shifted from the ground state values,
and depend on the mutual inductance. Numerical calculations show
that practically the same hysteretic behavior shown in
Fig.~\ref{doubling_theory} can be obtained for junctions A and B
both in the $0$ state with similar parameters of the double SQUID
\cite{stevan}.

To conclude, we studied an rf SQUID containing an inhomogeneous
Josephson junction as a weak link. We have found experimentally a
doubling of the switching frequency in certain ranges of magnetic
flux. The inhomogeneous junction can be modelled as a small dc SQUID
with $0$ and $\pi$ Josephson junctions. This model explains the
observed switching anomaly by a bistable switching of the
orientation of the current in the weaker section of the junction.
The suggested mechanism is effective, independent of the shape of
the CPR in both junctions, and may also be relevant for similar
observations in other systems.

We thank M.~Reinwald for help with the preparation of the
GaAs/AlGaAs-heterostructures. This work has been supported by the
German Science Foundation within SFB 689, the Serbian Ministry of
Science, Project No. 141014, the  Franco-Serbian PAI EGIDE Project
No. 11049XG, and US DOE project MA-509-MACA.


\end{document}